\documentclass[twocolumns]{IEEEtran}
\usepackage{spconf,amsmath,epsfig}
\usepackage{amssymb,bm}

\def\bs{\mathbf{s}}
\def\bv{\mathbf{v}}
\def\bw{\mathbf{w}}
\def\bx{\mathbf{x}}
\def\by{\mathbf{y}}

\def\br{\mathbf{r}}
\def\bpsi{\bm{\psi}}

\title{Performance analysis of a cross-layer collaborative beamforming approach in the presence of channel
and phase errors}
\twoauthors
  {Lun Dong, Athina P. Petropulu}
   {Dept. of Electrical \& Computer Engineering\\
   Drexel University, Philadelphia, PA 19104}
  {H. Vincent Poor \thanks{This work was supported by the
National Science Foundation under Grants ANI-03-38807, CNS-06-25637
and CNS-04-35052, and by the Office of Naval Research under Grant
ONR-N00014-07-1-0500.}}
   {School of Engineering \& Applied Science \\
   Princeton University, Princeton, NJ 08544}
\begin{document}

\maketitle
\begin{abstract}
Collaborative beamforming enables nodes in a wireless network
 to transmit a common message over long distances in an
energy efficient fashion. However, the process of making available
the same message to all collaborating nodes introduces delays. The
authors recently proposed a MAC-PHY cross-layer scheme that enables
collaborative beamforming with significantly reduced collaboration
overhead. The method requires knowledge of node locations and internode channel
coefficients. In this paper,  the performance of that approach is
studied analytically in terms of average beampattern and symbol
error probability (SEP) under realistic conditions, i.e., when
imperfect channel estimates are used and when there are phase errors
in the contributions of the collaborating nodes at the receiver.
\end{abstract}
\section{Introduction}
\label{intro}

Distributed, or collaborative, beamforming has been of considerable recent
interest as a preferred solution for long-distance transmission in
wireless networks, due to its energy efficiency
\cite{Mudumbai:2007},\cite{Ochiai}. In conventional distributed
beamforming schemes, a set of distributed nodes (called
collaborating nodes) act as a virtual antenna array and form a beam
to cooperatively transmit a \emph{common}  signal arising from a source node.
Using knowledge of network coordinates, each collaborating node
adjusts its initial phase so that the resulting beampattern focuses in the
direction of the desired destination. The requirement
that all collaborating nodes have access to the same message signal
means that source nodes must share their message signals with
collaborating nodes before beamforming. To study network
performance, one must take into account the overhead
(information-sharing time) required for node collaboration. If a
time-division multiple-access (TDMA) scheme were to be employed, the
information-sharing time would increase proportionally to the number
of source nodes.

The authors recently proposed a MAC-PHY cross-layer technique in
\cite{GLOBECOM:2007} and \cite{CISS:2007}, based on the idea of
collaborative beamforming of \cite{Ochiai}, to reduce the time
required for information sharing and to allow simultaneous multiple
beams. The main idea is as follows: for information-sharing, we
consider a real physical model in which collaborating nodes receive
linear mixtures of transmitted packets. Subsequently, each
collaborating node transmits a weighted version of its received
signal. The weights allow packets bound to the same destination to
add coherently at the destination node. Each collaborating node
computes its weight based on the estimated channel coefficients
between sources and itself, and also based on estimates of node
coordinates. In \cite{GLOBECOM:2007} and \cite{CISS:2007} the
analysis was performed under the assumption that all required
estimates are perfect. In this paper, we investigate performance
under imperfect conditions, i.e., when there are channel estimation
errors and phase errors.

\section{System model and proposed scheme} \label{proposed}
\begin{figure} [htp]
\centering \centerline{\epsfig{figure=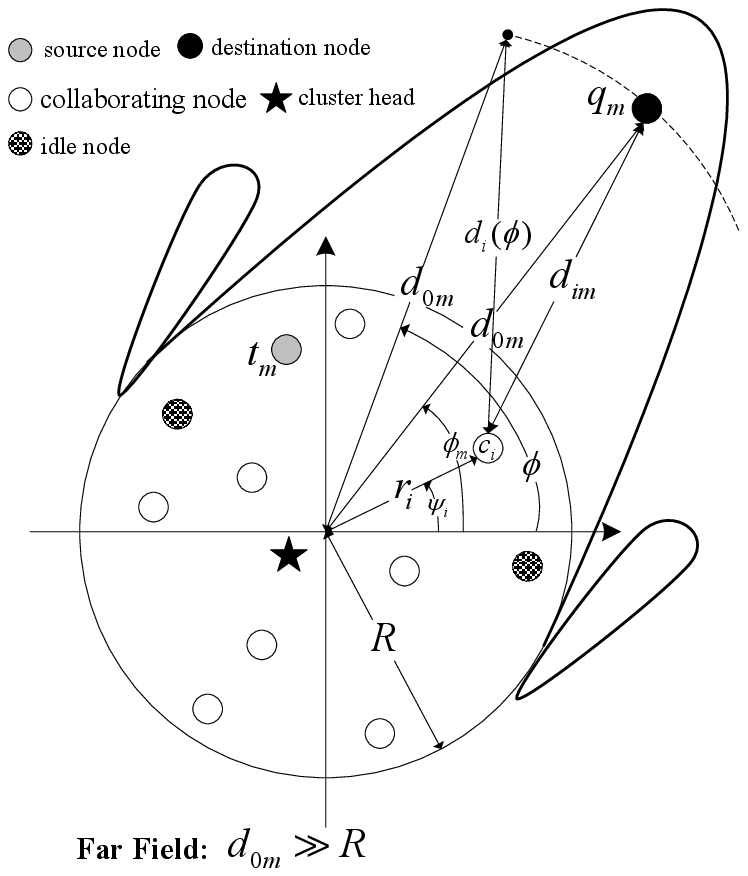,width=5cm}}
\caption{Illustration of notation.} \label{illustration}
\end{figure}

 The notation used here is
illustrated in Fig. \ref{illustration}. For simplicity, let us
assume that sources and destinations are coplanar. The network is
divided into clusters, so that nodes in a cluster can hear each
other's transmissions. During slot $n$, source nodes $t_1,\ldots, t_K$ in cluster
$C$ tend to communicate with nodes $q_1,\ldots,q_K$ that belong to
clusters $C_1,\ldots,C_K$, respectively. The beamforming is
performed by nodes in cluster $C$. The  $N$ collaborating
nodes, designated as $c_1,\ldots,c_N$, are assumed to be uniformly distributed
over a disk of radius $R$. We denote the location of $c_i$ in polar
coordinates with respect to the origin of the disk by
$(r_i,\psi_i)$. Let $d_{im}$ represent the distance between $c_i$
and the destination $q_m$, and $d_{0m}$ represent the distance
between the origin of the disk and $q_m$. If $\phi_m$ is the
azimuthal angle of $q_m$ with respect to the origin of the disk, the
polar coordinates of $q_m$ are $(d_{0m},\phi_m)$. Moreover, let
$d_{i}(\phi)$ denote the distance between $c_i$ and some receiving
point with polar coordinate $(d_{0m}, \phi)$.

We further make the following assumptions: (1) A slotted packet
system is considered, in which each packet requires one slot for its
transmission. Perfect synchronization is assumed between nodes in
the same cluster. Nodes operate under half-duplex mode, i.e., they
cannot receive while they are transmitting. (2) Nodes transmit
packets consisting of phase-shift keying (PSK) symbols each having
the same power $\sigma_s^2$. (3) Communication takes place over flat
fading channels. The channel gain during slot $n$ between source
$t_i$ and collaborating node $c_j$ is denoted by $a_{ij}(n)$. For
intra-cluster communications small-scale fading plays the dominant
role. Thus, for a fixed $n$, we model $a_{ij}(n)$ as circularly
symmetric complex Gaussian random variables with zero means and
variances $\sigma_a^2$ (i.e., this is a Rayleigh fading model). The gains of
different paths are assumed to be independent and identically distributed (i.i.d.).
The gain of a given path is constant during the slot duration.
(4) For inter-cluster communications, large-scale fading plays the
dominant role. We assume that the distances between collaborating
nodes and destinations are much greater than the maximum distance
between source and collaborating nodes. Thus, the complex
baseband-equivalent channel gain between nodes $c_i$ and $q_m$
during beamforming equals $ b_{m} e^{{j\frac{2\pi}{\lambda}}d_{im}}$
\cite{tse-book}, where $\lambda$ is the signal wavelength and
$b_{m}$ is the path loss between the center of the disk containing
the collaborating nodes and the destination.

In slot $n$, all source nodes within the cluster $C$ simultaneously
transmit their packets. The packet transmitted by node $t_j$
consists of $L$ symbols $\bs_j(n) \triangleq [s_j(n;0), \ldots,
s_j(n;L-1)]$. Due to the broadcast nature of the wireless channel,
non-active nodes in cluster $C$ hear a collision, i.e., a linear
combination of the transmitted symbols. More specifically, node
$c_i$ hears the signal
\begin{equation} \label{crec}
\bx_i(n)=\sum_{j=1}^K a_{ji}(n)\bs_j(n) + \bw_i(n)
\end{equation}
where $\bw_i(n)=[w_i(n;0),\ldots, w_i(n;L-1)]$ represents noise
 at the receiving node $c_i$.
 The noise is assumed to be zero-mean with covariance matrix
 $\sigma_w^2{\bf I}_L$, where ${\bf I}_L$ denotes the $L \times L$ identity matrix.

Suppose that in  slot $n+m, \ m=1,\ldots ,K$, the collaborating
nodes need to beamform $\bs_m(n)$ to destination $q_m$. Each
collaborating node $c_i$ transmits the signal
\begin{equation}
 \tilde \bx_i(n+m)=\bx_i(n) \mu_m a^*_{mi}(n) e^{-j \frac{2\pi}{\lambda}d_{im}}
\end{equation}
where $e^{-j \frac{2\pi}{\lambda}d_{im}}$ is the initial phase of
$c_i$. $\mu_m$ is a scalar used to adjust the transmit power; it is
the same for all collaborating nodes, and is on the order of $1/N$.

Given the collaborating nodes at radial coordinates
${\br}=[r_1,...,r_N]$ and azimuthal coordinates
 ${\bpsi}=[\psi_1,...,\psi_N]$, the
received signal at an arbitrary location with polar coordinates
$(d_{0m},\phi)$, is
\begin{equation} \label{rec}
\by(\phi;m|{\br,\bpsi})= \sum_{i=1}^N b_m \tilde \bx_i(n+m) e^{j
\frac{2\pi}{\lambda}d_i(\phi)} + \bv(n+m)
\end{equation}
where $\bv(n+m)$ represents noise at the receiver during slot $n+m$.
The covariance matrix of $\bv(n+m)$ equals
  $\sigma_v^2{\bf I}_L$.

The received signal at the destination $q_m$  during slot $n+m$ is
\begin{eqnarray}
\by(\phi_m;m|{\br,\bpsi})&=& \sum_{i=1}^N b_m \tilde \bx_i(n+m) +
\bv_m(n+m) \ .
\end{eqnarray}

It was shown in \cite{CISS:2007} that, as $N\rightarrow \infty$ and
omitting the noise, $\by(\phi_m;m|{\bf r,\bpsi}) \rightarrow N \mu_m
b_m \sigma_a^2 \bs_m(n)$. Thus, the destination node $q_m$ receives
a scaled version of $\bs_m(n)$. The beamforming step is completed in
$K$ slots, reinforcing one source signal at a time. Compared with
the scheme in \cite{Ochiai}, the information sharing time is reduced
from $K$ to 1. Multiple beams can be formed in one slot when source
packets have distinct destinations. In the rest of the paper, for
simplicity we will consider only the case in which a single beam is
formed during slot $n+m$, focusing on destination $q_m$.

\section{Average Beampattern under Imperfect Conditions} \label{beampattern}
The beampattern represents the distribution of received power along all
azimuthal angles. We showed in \cite{CISS:2007} that, under perfect
conditions the average beampattern is of a form similar to
\cite{Ochiai}, with increased sidelobe level. In this section, we
discuss effects of imperfect channels and phase on average
beampattern, respectively.

\subsection{Imperfect Channels} \label{imperfectCHbeam}
We model $\hat{a}_{mi}=a_{mi}+\delta a_{mi}$ as the imperfect
estimate of $a_{mi}$, where $\delta a_{mi}$ is the estimation error.
The estimation errors are i.i.d. Gaussian random variables, $\delta a_{mi}
\sim \mathcal{CN}(0, \sigma_{\delta}^2)$. The average beampattern
with imperfect channels can be expressed as
\begin{equation}
\tilde{P}_{\mathrm{av}}(\phi)= E\{|y(\phi; m|{\br,\bpsi})|^2\} =
P_{\mathrm{av}}(\phi)+\delta P_{\mathrm{av}}(\phi)
\end{equation}
where $P_{\mathrm{av}}(\phi)$ is the average beampattern related to
perfect channels $a_{mi}$, and $\delta P_{\mathrm{av}}(\phi)$ is the
average beampattern related to the estimation error $\delta a_{mi}$.
Following steps similar to those leading to $P_{\mathrm{av}}(\phi)$
in \cite{CISS:2007}, one can obtain
\begin{eqnarray}
\delta P_{\mathrm{av}}(\phi)= \mu_m^2 b_m^2E \left\{ |s_m|^2
\sum_{i=1}^N |a_{mi}|^2 |\delta a_{mi}|^2 \right. \nonumber\\
 + \sum_{j=1 \atop j \neq m}^{K} |s_j|^2 \sum_{i=1}^{N}|\delta
a_{mi}|^2
|a_{ji}|^2 \nonumber +\left. \sum_{i=1}^N {|\delta a_{mi}|^2}|w_i|^2 \right\} \nonumber \\
=N^2 \mu_m^2 b_m^2\sigma_s^2\sigma_a^4\left( \frac{K
\sigma_\delta^2}{N \sigma_a^2 }+\frac{\sigma_\delta^2}{N \gamma_1
\sigma_a^2}\right) \propto \frac{\sigma_\delta^2}{\sigma_a^2}
\end{eqnarray}
where $\gamma_1 \buildrel \triangle \over=
\sigma_s^2\sigma_a^2/\sigma_w^2$ represents the average SNR at the
collaborating nodes. Note that $\delta P_{\mathrm{av}}(\phi)$ is
actually a constant independent of $\phi$. In other words, the
effect of imperfect channels on average beampattern is an increased
sidelobe level.

\subsection{Imperfect Phase} \label{imperfectPHbeam}

Under imperfect phase, each collaborating node $c_i$ will transmit
the signal $\tilde \bx_i(n+m) e^{j \tau_i}$, where the $\tau_i$
represents the phase error,  which is assumed i.i.d. with respect to $i$. We use
the same model as in \cite{Ochiai} for the phase errors. Regarding
how to obtain the initial phase, two cases (closed-loop and
open-loop) are considered (see \cite{Ochiai} for details):

(1) For the closed-loop case, imperfect phase corresponds to the
phase offset due to the phase ambiguity caused by carrier phase
jitter or offset between the transmitter and receiver nodes. We
assume that the phase error $\tau$ follows a Tikhonov distribution,
which is a typical phase jitter model for phase-locked loop (PLL)
circuits.

(2) For the open-loop case, imperfect phase results from estimation
errors in the location parameters $r_i$ and $\psi_i$. We assume the
corresponding radius error $\delta r_i$ is uniformly distributed
over $[-r_{max}, r_{max}]$, and the angle error $\delta \psi_i$ is
uniformly distributed over $[-\psi_{max}, \psi_{max}]$. The radius
and angle errors are further assumed to be mutually independent random
variables, independent of $r_i$ and $\psi_i$.

Based on the above phase error models, we can show that the
expressions of the average beampattern are similar to the results in
Section VI of \cite{Ochiai} with the only difference being a scaling
factor. Thus, as in \cite{Ochiai}, the basic effect of these phase
errors is in reducing the power in the main lobe. The derivation is
similar to that in \cite{Ochiai} and is omitted here due to space
limitations.

\section{SEP under Imperfect Conditions} \label{sec:SEP}
Under perfect conditions, the received signal (one sample) at the
destination is given by
\begin{eqnarray}
y(\phi_m;m)= \mu_m b_m\sum_{i=1}^N |a_{mi}|^2 s_{m}+ \mu_m
b_m\sum_{i=1}^N a_{mi}^* \eta_i +v \ ,
\end{eqnarray}
where $\eta_i \triangleq \sum_{j=1 \atop j\neq m}^{K}{a_{ji}
s_{j}}+w_{i}$. We showed in \cite{GLOBECOM:2007} that  $ \eta_i \sim
\mathcal{CN}\left(0, \sigma_\eta^2\right)$ where
$\sigma_\eta^2\triangleq(K-1)\sigma_a^2\sigma_s^2+\sigma_w^2$.

Given $a_{mi}$, the instantaneous signal-to-interference-plus-noise ratio (SINR), $\gamma$, equals
\begin{equation} \label{SNR}
\gamma= \frac{\mu_m^2 b_m^2 \sigma_s^2 \xi^2}{\mu_m^2 b_m^2
\sigma_\eta^2 \xi +\sigma_v^2}
\end{equation}
where $\xi \buildrel \triangle \over = \sum_{i=1}^N |a_{mi}|^2$
follows an Erlang distribution ($\xi \sim \mathrm{Erlang} (N,
\sigma_a^2) $).

Given $K$, the SEP for M-PSK symbols under perfect conditions is
\cite{GLOBECOM:2007}, \cite{simon-book}
\begin{eqnarray} \label{SEP}
P_s(K) &=& \frac{1}{\pi}\int_{0}^{\frac{(M-1)\pi}{M}} \int_0^\infty
\exp \left(-\frac{\sin^2(\pi/M)}{\sin^2\varphi} \cdot \gamma \right)
\nonumber\\
&& \times \frac{\xi^{N-1}e^{-\frac{\xi}{\sigma_a^2}}}{\sigma_a^{2N}
(N-1)!} d \xi d \varphi \ .
\end{eqnarray}

\subsection{Imperfect Channels}
Taking channel errors into account, the received signal at the
destination is given by
\begin{eqnarray}
y(\phi_m;m)&=& \mu_m b_m\sum_{i=1}^N |a_{mi}|^2 s_{m} + \mu_m
b_m\sum_{i=1}^N a_{mi} \delta a_{mi}^* s_{m}\nonumber\\ &&+ \mu_m
b_m\sum_{i=1}^N (a_{mi}^*+\delta a_{mi}^*) \eta_i +v \ .
\end{eqnarray}

Since the destination node does not have  knowledge of $\delta
a_{mi}$, the term $\mu_m b_m\sum_{i=1}^N a_{mi} \delta a_{mi}^*
s_{m}$ represents interference. Thus, in the interference term,
$a_{mi}$ and $\delta a_{mi}$ are coupled together, and the exact SEP
would involve integration of all of the $2N$ random variables
($a_{mi}$ and $\delta a_{mi}$, $i=1,\ldots,N$). In the sequel we
will use an approximation that simplifies this analysis.

Let us define
\begin{equation}
\kappa = \mu_m b_m\sum_{i=1}^N \left[ a_{mi} \delta a_{mi}^* s_{m} +
(a_{mi}^*+\delta a_{mi}^*) \eta_i \right] \ .
\end{equation}
It is easy to show that, given $a_{mi}$, $E\{\kappa\} = 0$ and
\begin{equation}
\sigma_{\kappa}^2 = E\{|\kappa|^2\} =  \mu_m^2 b_m^2
(\sigma_\eta^2+\sigma_s^2 \sigma_\delta^2)\xi + \mu_m^2 b_m^2
\sigma_\eta^2 N\sigma_\delta^2 \ .
\end{equation}
According to the central limit theorem, when $N$ is large, $\kappa$ is
approximately normally distributed.
Let us thus approximate the distribution of  $\kappa$ as $\kappa
\sim \mathcal{CN}(0, \sigma_{\kappa}^2)$. Taking into account the independence of
$\kappa$ and $v$, the
approximate instantaneous SINR, $\gamma_{\mathrm{ch}}$, equals
\begin{eqnarray} \label{approxSEPcherr}
\gamma_{\mathrm{ch}} & = &  \frac{\mu_m^2 b_m^2 \sigma_s^2
\xi^2}{\mu_m^2 b_m^2 (\sigma_\eta^2+\sigma_s^2 \sigma_\delta^2)\xi +
\mu_m^2 b_m^2 \sigma_\eta^2 N\sigma_\delta^2+\sigma_v^2}
\end{eqnarray}
which contains only a single random variable $\xi$. Finally, to
calculate the SEP under imperfect channel conditions, let us
substitute $\gamma_{\mathrm{ch}}$ for $\gamma$ in (\ref{SEP}). The techniques in
section IV-A of \cite{GLOBECOM:2007} can be used to obtain simple
bounds for SEP.

\begin{figure} [htp]
\centering \centerline{\epsfig{figure=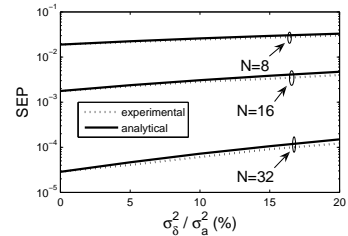,width=4.4cm}}
\caption{SEP under imperfect channels.  $K=4$, $\gamma_1=20$ dB,
$\gamma_2 \triangleq N^2 \mu_m^2 b_m^2 \sigma_s^2
\sigma_a^4/\sigma_v^2 = 20$ dB, BPSK symbols.} \label{channelerr}
\end{figure}

\underline{\textbf{Simulation}}: Fig. \ref{channelerr} shows the SEP versus
$\sigma_\delta^2/\sigma_a^2$. As expected, SEP increases with increasing
$\sigma_\delta^2$. The analytical result based on
(\ref{approxSEPcherr}) matches well experimental results for a wide
range of values of $\sigma_\delta^2$.

\subsection{Imperfect Phase}

Imperfect phase has two effects on the receiver: signal power
reduction and phase distortion \cite{Mudumbai:2007}. Assuming that the phase
distortion is compensated for by the coherent receiver (e.g., by
pilots), here we focus on signal power reduction only.

We define the power reduction coefficient $A_{\tau}=
P_{\mathrm{err}}/P_{\mathrm{ideal}} \leq 1$, where
$P_{\mathrm{err}}$ is the average received signal power with phase
error and $P_{\mathrm{ideal}}$ is the average signal power under
perfect phase.

Taking phase errors into account, the received signal at the
destination is given by
\begin{eqnarray}
y(\phi_m;m)&=& \mu_m b_m\sum_{i=1}^N |a_{mi}|^2 e^{j \tau_i}
s_{m} \nonumber\\
&&+ \mu_m b_m\sum_{i=1}^N a_{mi}^* \eta_i e^{j \tau_i} +v \ ,
\end{eqnarray}
where $\tau_i$ is the phase error of collaborating node $c_i$.

Note that the statistics of $\eta_i e^{j \tau_i}$ are the same as those of
$\eta_i$; so phase errors do not change the statistical behavior of the
interference term.

Assuming we use a coherent receiver, the instantaneous
SINR, $\gamma_{\mathrm{ph}}$, equals
\begin{equation}
\gamma_{\mathrm{ph}}= \frac{\mu_m^2 b_m^2  \sigma_s^2 |\sum_{i=1}^N
|a_{mi}|^2 e^{j \tau_i} |^2}{\mu_m^2 b_m^2 \sigma_\eta^2 \xi
+\sigma_v^2} \ .
\end{equation}
Since $|a_{mi}|^2$ and $e^{j \tau_i}$ are coupled, the
exact SEP  includes  integration with respect to $a_{mi}$ and
$\tau_i$ ($i=1,\ldots,N$), which is computationally complex. To
facilitate analysis we make the following approximation for
$\gamma_{\mathrm{ph}}$:
\begin{eqnarray} \label{approxSEPpherr}
\gamma_{\mathrm{ph}} \approx A_{\tau} \cdot \frac{\mu_m^2 b_m^2
\sigma_s^2 (\sum_{i=1}^N |a_{mi}|^2)^2}{\mu_m^2 b_m^2 \sigma_\eta^2
\xi +\sigma_v^2}  = A_{\tau} \cdot \gamma \ .
\end{eqnarray}
In other words, $\gamma_{\mathrm{ph}}$ is approximated by the
instantaneous SINR under perfect conditions scaled down by a
coefficient $A_{\tau}$.

It can be shown that
\begin{eqnarray} \label{Atau}
A_{\tau}= \frac{P_{\mathrm{err}}}{P_{\mathrm{ideal}}} =
\frac{2+(N-1)\left| E \left\{ e^{j \tau_i}\right \}\right| ^2}{N+1}
\ ,
\end{eqnarray}
where $\left| E \left\{ e^{j \tau_i}\right \}\right| ^2$ depends on
the specific phase error model used. Based on the phase error models
in section \ref{imperfectPHbeam}, $\left| E \left\{ e^{j
\tau_i}\right \}\right| ^2$ has been derived in \cite{Ochiai}.

\begin{figure}[htb]
\begin{minipage}[b]{.48\linewidth}
  \centering
 \centerline{\epsfig{figure=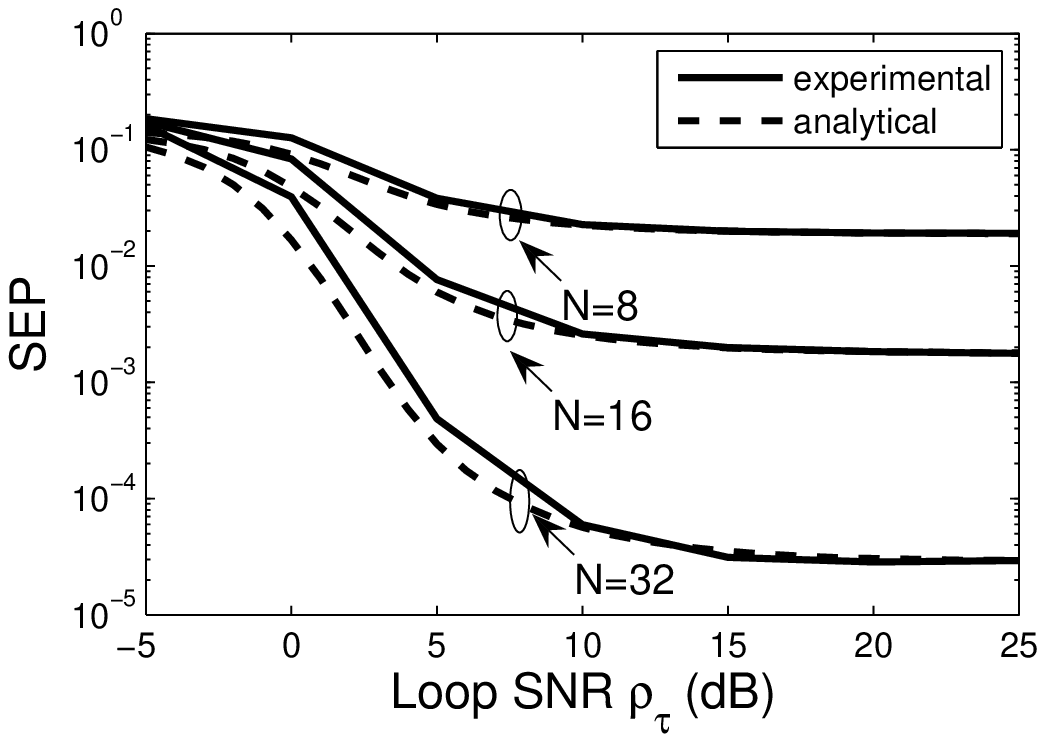,width=4.0cm}}
  \centerline{(a) closed-loop}
\end{minipage}
\hfill
\begin{minipage}[b]{0.48\linewidth}
  \centering
 \centerline{\epsfig{figure=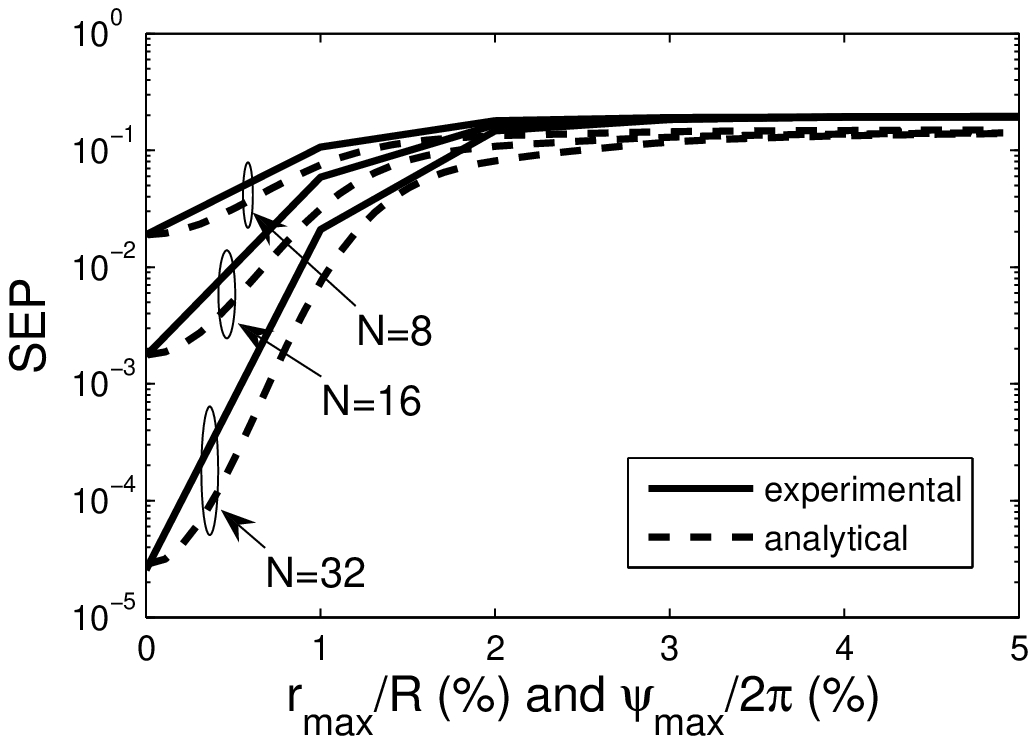,width=3.8cm}}
  \centerline{(b) open-loop}
\end{minipage}
\caption{SEP under imperfect phase. $R=10\lambda$, $K=4$,
$\gamma_1=20$ dB, $\gamma_2 \triangleq N^2 \mu_m^2 b_m^2 \sigma_s^2
\sigma_a^4/\sigma_v^2 = 20$ dB, BPSK symbols.} \label{phaserr}
\end{figure}

\underline{\textbf{Simulation}}: In Fig. \ref{phaserr}(a), we show
the SEP as a function of loop SNR $\rho_{\tau}$ (closed-loop case).
The variance of  the phase error is $1/\rho_{\tau}$. For the
analytical SEP, we directly use the results in \cite{Ochiai} to
calculate $\left| E \left\{ e^{j \tau_i}\right \}\right| ^2$ and
obtain $A_\tau$ in (\ref{Atau}). As observed, $\rho_{\tau}>10$ dB
may be necessary to achieve a satisfactory SEP. Fig.
\ref{phaserr}(b) shows the SEP vs. $r_{max}/R$ and
$\psi_{max}/(2\pi)$ (open-loop case), where both radius and angle
errors are considered. One can see that the phase error in the
open-loop case can severely degrade the SEP performance. Thus, it is
important to investigate techniques that enable accurate location
estimation. In both figures, the analytical result based on
(\ref{approxSEPpherr}) match well the experimental results for a
wide range of phase errors.

\section{conclusions} \label{con}
We have considered the cross-layer collaborative beamforming
approach of \cite{GLOBECOM:2007} and  \cite{CISS:2007}, and we have analyzed its
performance under imperfect conditions. For the average beampattern,
the principal effect of imperfect channel information is increased sidelobe
level, and the principal effect of imperfect phase information is reduced
mainlobe power. We have provided
approximate analytical expressions for the SEP under imperfect conditions,
which show the effects of imperfect channel and phase on that quantity.

\vspace{-0.1in}

\end{document}